\documentclass{nature}

\usepackage{color}
\usepackage{graphicx}
\usepackage{verbatim}
\usepackage{xcolor}
\usepackage{amsmath}
\usepackage{upgreek}

\makeatletter
\let\saved@includegraphics\includegraphics
\AtBeginDocument{\let\includegraphics\saved@includegraphics}
\renewenvironment*{figure}{\@float{figure}}{\end@float}
\makeatother

\title{Spin-Orbit-Torque Driven Propagating Spin Waves}

\author{H. Fulara$^1$, M. Zahedinejad$^{1,2}$, R. Khymyn$^1$, A. A. Awad$^{1,2}$, S. Muralidhar$^1$, M. Dvornik$^{1,2}$ \& J. \AA kerman$^{1,2,3}$}

\begin{document}

\maketitle

\begin{affiliations}
  \item Physics Department, University of Gothenburg, 412 96 Gothenburg, Sweden
 \item NanOsc AB, Electrum 229, 164 40 Kista, Sweden
 \item Material and Nanophysics, School of Engineering Sciences, KTH Royal Institute of Technology, Electrum 229, 164 40 Kista, Sweden
\end{affiliations}

\begin{abstract}
Spin-orbit torque (SOT) can drive sustained spin wave (SW) auto-oscillations in a class of emerging microwave devices known as spin Hall nano-oscillators (SHNOs), which have highly non-linear properties governing robust mutual synchronization at frequencies directly amenable to high-speed neuromorphic computing. However, all demonstrations have relied on localized SW modes interacting through dipolar coupling and/or direct exchange. As nanomagnonics requires propagating SWs %as nanoscale signal carriers 
for data transfer, and additional computational functionality can be achieved using SW interference, SOT driven propagating SWs would be highly advantageous. 
Here, we demonstrate how perpendicular magnetic anisotropy can raise the frequency of SOT driven auto-oscillations in magnetic nano-constrictions well above the SW gap, resulting in the efficient generation of field and current tunable propagating SWs. Our demonstration greatly extends the functionality and design freedom of SHNOs enabling long range SOT driven SW propagation for nanomagnonics, SW logic, and neuromorphic computing, %based on oscillator networks 
directly compatible with CMOS technology. 
\end{abstract}

\section*{Introduction}

The recent emergence of spin-orbit torque (SOT)\cite{Miron2011nat,brataas2014natnano,gambardella2011philtrans} %, derived 
from pure spin currents, has opened a new avenue for the controlled manipulation of magnetic moments in spintronic devices resulting in dramatically improved efficiency\cite{liu2012science,mellnik2014nature} and much lower power dissipation\cite{liu2012prl,avci2016natmat} compared to %than the one with 
conventional spin-transfer torque (STT)\cite{ralph2008jmmm}. Thanks to the %unprecedented 
ability of SOT to compensate natural magnetic damping over spatially extended regions, there has been considerable interest in exploring the long range enhancement of spin wave (SW) propagation\cite{wang2011prl,padron2011apl,liu2018natcomm,demidov2014magnonicsapl,evelt2016apl,gladii2016apl,divinskiy2018arxiv} in a variety of nanoscale devices with the aim of developing an energy-efficient and ultra-high-speed beyond-CMOS spin-wave-based technology for signal processing\cite{chumak2014ntcom,balinskiy2018aip} and computation\cite{chumak2015ntmat,bracher2018jap,liu2018natcomm,chumak2019arXiv}, so-called magnonics\cite{neusser2009adv,kruglyak2010iop}. 

One of the most promising SOT devices for active, controllable SW generation on the nano-scale is the nano-constriction based spin Hall nano-oscillator (SHNO)\cite{Demidov2014apl,Kendziorczyk2016prb,Awad2016natphys,durrenfeld2017nanoscale,Chen2016procieee,divinskiy2017apl}. It can be easily fabricated using a wide range of different bilayer combinations\cite{Zahedinejad2018apl,mazraati2016apl,evelt2018pra,divinskiy2017apl} and the generated SWs can be directly observed using both electrical and optical microwave spectroscopy\cite{Zahedinejad2018apl,mazraati2016apl,evelt2018pra,divinskiy2017apl,Awad2016natphys,mazraati2018pra,mazraati2018arxiv}. Most importantly for applications, nano-constriction SHNOs exhibit highly robust mutual synchronization, both in long chains\cite{Awad2016natphys} and in two-dimensional arrays\cite{Zahedinejad2018arxiv}, which both improves their signal properties by orders of magnitude and lend themselves to neuromorphic computing\cite{romera2018nature,grollier2016ieee}.

A major limitation of nano-constriction SHNOs, however, is the localized nature of the SOT driven auto-oscillations.\cite{Dvornik2018prappl} The localization is a consequence of the easy-plane anisotropy and the geometry of the device, which lead to a negative magnetodynamic non-linearity, further exacerbated by the Oersted field from the drive current, and from the SOT itself.\cite{Dvornik2018prappl,Dvornik2018arxiv} It would be highly advantageous if the localization could be mitigated so as to generate truly propagating SWs. Not only should this lead to mutual synchronization over much longer distances, it would also make SOT driven SWs directly applicable to additional non-conventional computing schemes such as wave based computing\cite{lenk2011pr,klingler2014apl,chumak2015ntmat,fischer2017apl}.

In a recent work\cite{evelt2018pra}, Evelt and coworkers demonstrated SOT driven propagating SWs in extended Bi-substituted YIG films with perpendicular magnetic anisotropy\cite{Soumah2018natcomm} (PMA). While the auto-oscillations could only be observed optically and did not exhibit any frequency tunability via the drive current, the demonstration raises the question whether the addition of PMA to metal based nano-constriction SHNOs could potentially lead to SOT generated propagating SWs in more practical devices directly compatible with CMOS technology\cite{Zahedinejad2018apl}. Here we show, using 150 nm and 200 nm nano-constrictions in W/CoFeB/MgO material stacks with substantial PMA, that it is indeed possible to generate strongly current-tunable propagating SWs over a very wide frequency range of about 3--23 GHz. The SWs are studied using electrical microwave spectroscopy and modelled using micromagnetic simulations. Auto-oscillations are observed at currents as low as 0.15 mA, where they are still localized and exhibit negative non-linearity. As the current is increased, the non-linearity changes sign and the localized SWs exhibit %transition smoothly 
smooth transition into propagating SWs at about 0.5 mA. It is hence possible to seamlessly turn on and off the localization, which will allow %for
the generation of ultra-short SW pulses driven by %using only 
the current alone, which 
is much faster than using external fields%in contrast to SW pulse generation using spin torque nano-oscillators and external fields
\cite{divinskiy2016apl}.

\subsection{Perpendicular magnetic anisotropy controlling the magnetodynamic non-linearity.}

The rich non-linear magneto-dynamics in patterned magnetic thin films can be analytically described by a single non-linearity coefficient, $\mathcal{N}$, the magnitude and sign of which determine the strength and nature of magnon-magnon interactions, with positive and negative values signifying magnon repulsion and attraction, respectively.\cite{slavini2005approximate,slavin2009nonlinear,Dvornik2018prappl} As spin transfer torque and SOT can generate very high SW amplitudes, the sign and magnitude of $\mathcal{N}$ leads to dinstinctly different behavior of the auto-oscillations. A negative non-linearity %allows 
makes the auto-oscillation frequency %to drop 
decrease with amplitude, eventually moving it into the magnonic band gap, where it %results in 
first leads to SW self-localization, and can further promote the nucleation of magnetodynamical solitons such as SW bullets\cite{slavin2005prl,bonetti2010prl,dumas2013prl} in easy-plane magnetic films and magnetic droplets\cite{mohseni2013science,chung2018direct} in films with very large PMA. A large positive non-linearity, on the other hand, makes the auto-oscillation frequency \emph{increase} with amplitude to well above the ferromagnetic resonance (FMR) frequency leading to the propagation of spin waves with a finite real wave-vector. A prominent example is the so-called spin torque driven Slonczewski modes\cite{Slonczewski1999jmmm,madami2011natnano,houshang2018natcomm}, which can also form SW beams\cite{Hoefer2008prb} in oblique fields, particularly useful for mutual synchronization\cite{Houshang2015natnano}.

\begin{figure}
  \begin{center}
  \includegraphics[width=15cm]{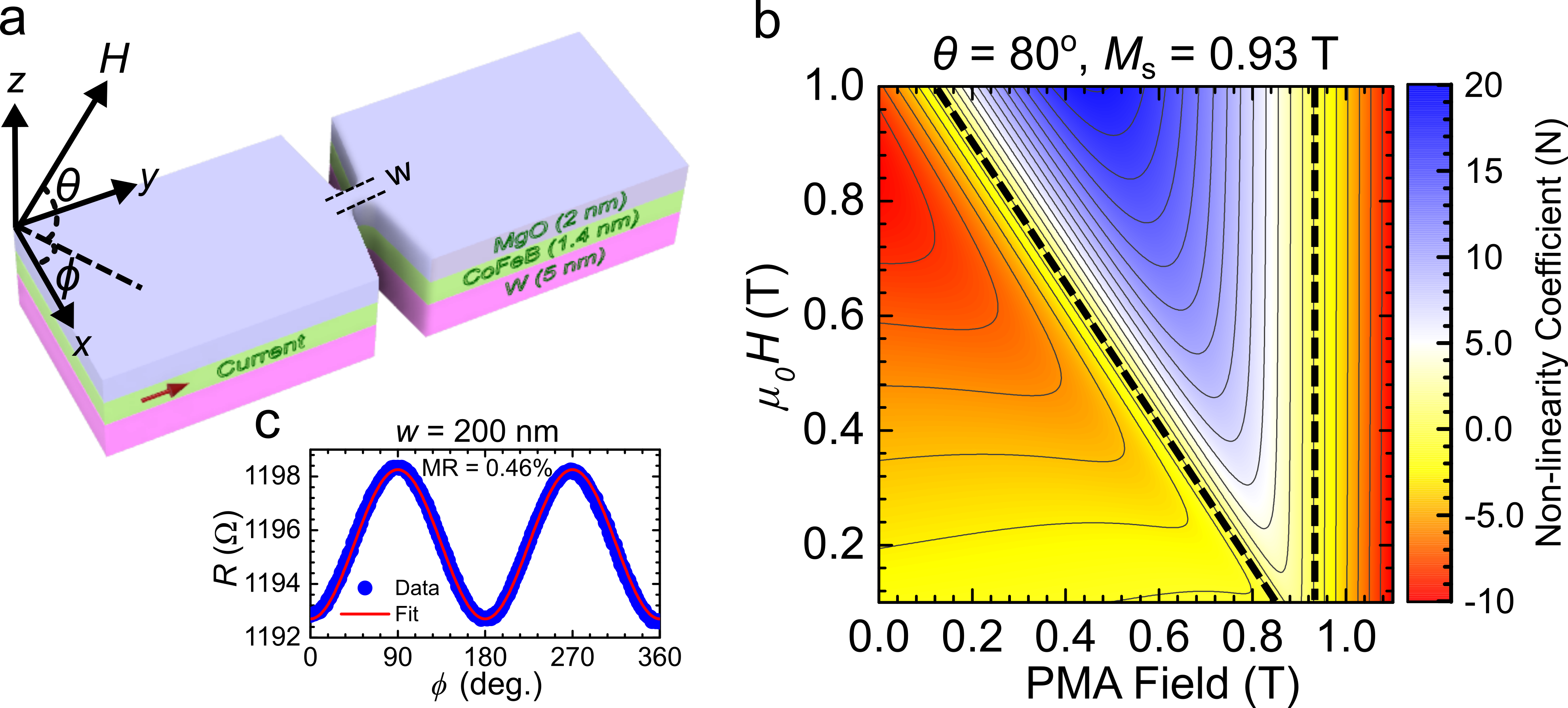}
    \caption{\textbf{Device schematic, magnetoresistance, and non-linearity coefficient.} (a) Schematic of a SHNO with nano-constriction width \textit{w}. (b) Contour plot displaying %the dependence of 
    the analytically calculated non-linearity coefficient ($\mathcal{N}$) as a function of PMA strength and applied out-of-plane ($\theta = 80^{\circ}$) field for a thin magnetic film with saturation magnetization, $\mu_0 M_{\text{S}} =$ 0.93 T; % of the material under investigation; 
    dashed black line indicates $\mathcal{N} = 0$. %points of zero non-linearity. 
    (c) Anisotropic magneto-resistance measured with 0.05 mA on a 200 nm nano-constriction under a rotating 70 mT in-plane field.}
    \label{fig:1}
    \end{center}
\end{figure}

The non-linearity is generally %defined 
governed by the applied field vector and/or an effective magnetic anisotropy tensor and %essentially attains a zero value 
is zero for an isotropic magnet regardless of the applied external field strength. The easy-plane %presence of an in-plane 
shape anisotropy of a magnetic thin film holds the magnetization vector in the film plane and therefore the non-linearity strongly depends on the strength and orientation of the applied field. However, anisotropy induced by %the film surface or 
an interface to a different material can counteract the shape anisotropy and even pull the magnetization vector out of the film plane. %The latter is known as 
Such PMA contributes a term with the opposite sign in the nonlinear coefficient compared to the %one with the 
shape anisotropy. %To gain more insight about the behaviour of non-linearity in magnetic thin films, we analytically calculated the magnitude and 
The impact of PMA on the sign and strength of $\mathcal{N}$ %the non-linearity coefficient 
can then be calculated for a thin magnetic film using the method given in Refs.~\citen{slavini2005approximate,slavin2009nonlinear} with the result plotted in Fig.~\ref{fig:1}b as a function of the strength of applied out-of-plane (OOP) field ($\theta_{\text{ex}}=80^{\circ}$) and the PMA field. % values consistent with the following study. 

One notes %can note from Fig. \ref{fig:1}b 
that $\mathcal{N}$ increases monotonically from negative (red regions) to positive (blue regions) values as a function of the OOP field strength and  goes through zero %while crossing over a point of zero non-linearity 
at a certain field value (black dashed line) that depends on the PMA strength\cite{bonetti2010experimental}, \emph{i.e.}~the PMA %induces an additional positive shift of $\mathcal{N}$ and therefore can 
shifts the point of zero non-linearity towards lower values of applied field. %Consequently, the magnetic layer becomes effectively isotropic with $N=0$ at lower fields, where the contribution of the PMA and the shape anisotropy to non-linearity are equal (see the right border at Fig.~\ref{fig:1}b)). 
Adding PMA hence makes it possible to reach positive non-linearity at much lower fields. As the auto-oscillation threshold current decreases with decreasing field, it should in principle be possible to drive propagating SWs at much lower currents, which the SHNO can sustain without degradation. 

As a side note, any further increase of the PMA beyond the point where it completely compensates the shape anisotropy, \emph{i.e.}~where the magnetization equilibrium angle changes from in-plane (IP) to perpendicular to the plane, again results in a %complete compensation of in-plane shape anisotropy of the film and the non-linearity 
negative $\mathcal{N}$ (the second red region at high PMA in Fig.~\ref{fig:1}b). % evidently turns into negative in OOP field configuration. 
As a consequence, the current dependence of the auto-oscillation frequency again becomes negative\cite{Rippard2010prb,Mohseni2011pssrrl}, the generated spin waves self-localize, and can eventually % \emph{e.g.}~ and has also been employed for the nucleation of self-localized 
nucleate magnetic droplet solitons\cite{mohseni2013science,Macia2014natnano,Lendinez2015prb,Lendinez2017prappl,Divinskiy2017prb,chung2018direct}. %in all-perpendicular nanocontact-based STNOs . 

\section*{Results}
\subsection{Nano-patterned SHNO device schematic and magneto-dynamics.}
\sloppy A schematic of a nano-constriction SHNO is shown in Fig.~\ref{fig:1}a. The material stack consisted of sputtered %patterned with a nano-constriction width \textit{w} on a 
$\beta$-W(5~nm)/Co$_{20}$Fe$_{60}$B$_{20}$(1.4~nm)/MgO(2~nm). %stack is shown in Fig.~\ref{fig:1}a. 
The $\beta$-phase of W has been shown to produce large SOT\cite{zhang2016apl,Demasius2016natcomm} and %the W/CoFeB and CoFeB/MgO interfaces add PMA to the 
a thinner CoFeB interfaced with MgO layer enhances PMA in the CoFeB layer\cite{ikeda2010natm}. %By tuning the CoFeB thickness to 1.4 nm we can hence add substantial PMA without pulling the CoFeB out-of-plane.
The stack was fabricated on highly resistive Si substrates to both dissipate the local heat generated during operation and to reduce microwave losses; the SHNOs are hence CMOS compatible\cite{Zahedinejad2018apl}. A positive direct current (d.c.) is injected from the signal pad to ground along the $y$-direction while $\phi$ and $\theta$ define the IP and OOP field angles, respectively. Fig.~\ref{fig:1}c shows the in-plane angular dependence of the anisotropic magneto-resitance (AMR) measured for a 200 nm wide nano-constriction SHNO exhibiting a relatively large overall AMR value of 0.46$\%$ between the parallel and perpendicular orientation.

\begin{figure}[h]
  \begin{center}
  \includegraphics[width=15cm]{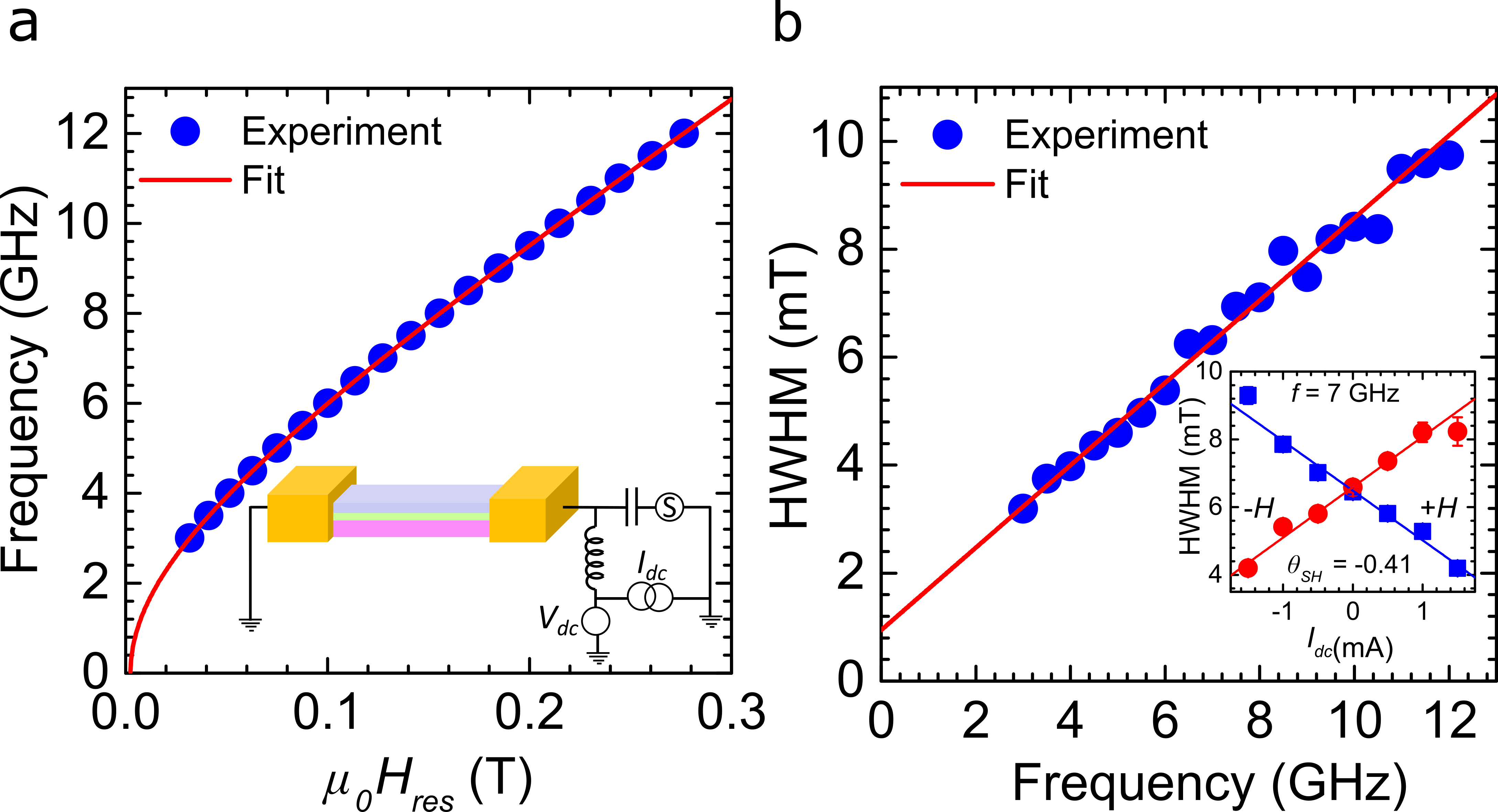}
    \caption{\textbf{ST-FMR measurements. % to extract Gilbert damping constant, PMA field, and spin-Hall angle.
    } (a) Resonance frequency vs.~in-plane field (blue dots) with a Kittel fit (red line) %Extracted resonance fields as a function of frequency (blue dots) and red solid line is a Kittel fit 
    yielding an effective magnetization $\mu_0M_{\text{eff}}= \mu_0(M_{\text{s}}-H_{\text{k}}^{\perp})$ of 0.31 T. %Using $M_{eff}= M_{S}-H_{k}^{\perp}$, we obtained a PMA field ($H_{k}^{\perp}$) of 0.62 T. 
    Inset: Illustration of the ST-FMR measurement on a 6$\times$18$~\mu m^2$ wide microstrip. (b) Extracted line-width (HWHM) vs.~resonance frequency yielding a Gilbert damping constant of $~\alpha =$ 0.023. Inset: Current dependent ST-FMR line-width for both positive (blue dots) and negative (red dots) field directions yielding a spin Hall angle of -0.41.}
    \label{fig:2}
    \end{center}
\end{figure}

Fig.~\ref{fig:2} summarizes the spin-torque ferromagnetic resonance (ST-FMR) measurements performed on a 6$\times$18$~\mu m^2$ microstrip of W/CoFeB/MgO to determine the %critical 
magneto-dynamical parameters. The inset of Fig.~\ref{fig:2}a schematically illustrates the experimental set-up employed for the ST-FMR (see Methods). % measurements and detailed in Methods subsection. 
The main panel of Fig.~\ref{fig:2}a shows the extracted resonance peak positions obtained at different microwave frequencies ranging from 3 to 12 GHz under RF current excitation. The resonance field dependence on frequency can be well fitted with the Kittel equation %and the best fit yields 
yielding an effective magnetization $\mu_0 M_{\text{eff}}=$ 0.31~T with a gyromagnetic ratio of $\gamma/2\pi$ = 29.9~GHz/T. With the saturation magnetization, $\mu_0 M_{\text{S}} =$ 0.93 T obtained from Alternating Gradient Magnetometry (AGM) measurements, we extract a PMA field of $\mu_0 H_{\text{k}}^{\perp}=\mu_0(M_{\text{S}}-M_{\text{eff}})=$ 0.62 T, \emph{i.e.}~while we have substantial PMA, it is not strong enough to pull the magnetization out-of-plane. % using equation $M_{\text{eff}}= M_{\text{s}}-H_{\text{k}}^{\perp}$\cite{liu2011jap}. 
Fig.~\ref{fig:2}b displays the plot of linewidths extracted as half width at half maximum (HWHMs) from the same resonance peaks as a function of different microwave frequencies and the linear best fit of experimental data gives rise to a %reasonably small 
Gilbert damping constant of $\alpha$ = 0.023. The inset of Fig.~\ref{fig:2}b shows the %d.c. 
current induced linewidth changes extracted at a fixed microwave frequency of 7 GHz for two opposite in-plane field orientations, the slope of which yields a high value of the spin-Hall angle (SHA), $\theta_{\text{SH}}=$ -0.41, typical for $\beta$-W and indicating the presence of large SOT from the spin-Hall effect in our devices\cite{Demasius2016natcomm}. 

\subsection{Propagating spin waves}
\begin{figure}[h]
  \begin{center}
  \includegraphics[width=16cm]{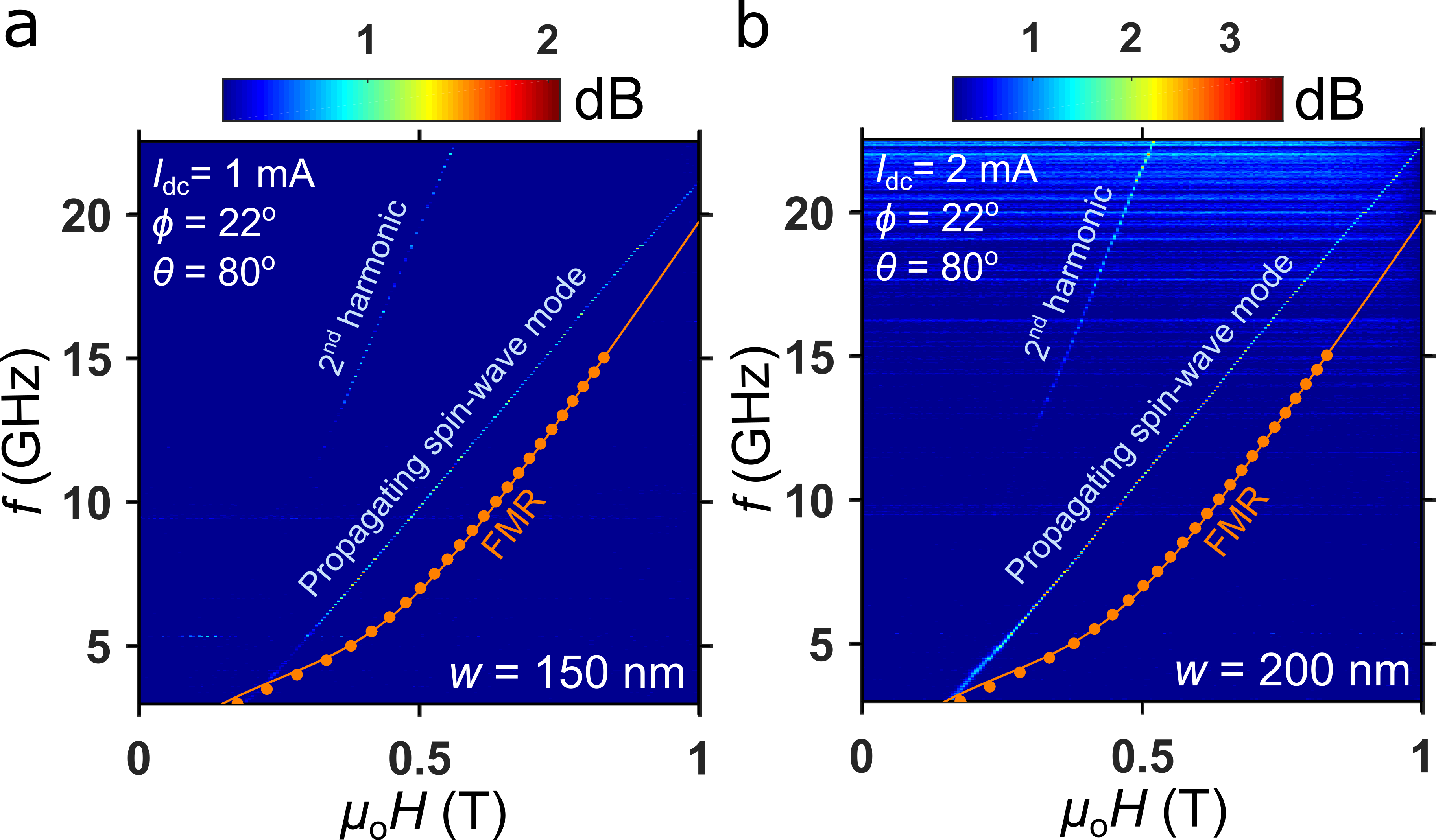}
    \caption{\textbf{Auto-oscillating %Generation of 
    propagating spin waves vs.~OOP field strength.%and confirmation with FMR spectrum extracted from ST-FMR measurements.
    } Power spectral densities (PSDs) vs.~%displaying field sweep spin-wave auto-oscillations excited in 
    out-of-plane field for a (a) 150 nm and (b) 200 nm nano-constriction. % width SHNO devices as a function of out-of-plane applied magnetic field strengths. 
    Orange data points are the ST-FMR resonances obtained under identical conditions on 4$\times$14$~\mu m^2$ wide microstrips %. % by employing ST-FMR measurements. 
    with the solid line being a fit to the Kittel equation (Eq. (\ref{eq:FMRfreq}) in Methods).% under magnetostatic boundary conditions.
    }
    \label{fig:3}
    \end{center}
\end{figure}

Figure \ref{fig:3} shows color plots of the generated microwave power spectral density (PSD) %of auto-oscillations 
as a function of OOP applied field strength measured for two different nano-constriction widths %SHNO devices excited 
with fixed direct currents of $I_{\text{dc}} =$ 1 and 2 mA, respectively. In both measurements, the IP field angle was fixed at $\phi$ = 22$^\circ$ to ensure sufficient electrical sensitivity to the auto-oscillation signal while the OOP field angle $\theta$ = 80$^\circ$ was chosen in a way to achieve large positive non-linearity in the active nano-constriction region. The orange circles, fitted with a solid orange line using the Kittel equation (Eq. (\ref{eq:FMRfreq})), % in Methods) under the magnetostatic boundary conditions, 
represent the FMR frequencies obtained from ST-FMR measurements on microstrips under identical applied field conditions. %configuration on a 4$\times$14$~\mu m^2$ wide microstrip. 
As the FMR frequency corresponds to %It may be noted that the FMR condition describes a spin-wave with infinitely large wavelength and therefore minimal 
a wave vector of $\vec{k}$ = 0, it allows us %. The FMR frequency therefore allows one 
to distinguish between %the excited 
propagating and localized spin-wave modes in the SHNO, since % nano-magnetic devices. While a 
spatially localized modes with a frequency well below FMR have no well-defined real wave vector, % and eventually fades out in the extended layer, 
while propagating modes with frequency higher than the FMR have a finite real $\vec{k}$. % and therefore can propagate efficiently within the ferromagnetic layer. 
%Fig.~\ref{fig:3}a shows the auto-oscillating mode behaviour of a 150 nm constriction %width SHNO device 
%excited with %continuous injection of 
%a constant direct current of 1 mA, exhibiting a linear rise in frequency at a rate of 22.8 MHz mT$^{-1}$ with increasing applied field strength. 

It is noteworthy that in all previously investigated nano-constriction and nano-gap SHNOs\cite{Demidov2012b,Liu2013,durrenfeld2017nanoscale}, the auto-oscillations remained localized with frequency lower than the FMR spectrum of the magnetic material. As can be seen in Fig.~\ref{fig:3}a, this is in our case only true for $\mu_0 H<$ 0.2 T and at all higher fields, the auto-oscillation frequency lies up to several GHz above the FMR frequency. This general behavior was observed in all devices as \emph{e.g.}~in %an additional example we show in 
Fig.~\ref{fig:3}b where a larger nano-constriction with $w$ = 200 nm follows the same trend, but now with improved microwave characteristics. It can be noted that in addition to a higher output power, the auto-oscillations in the larger nano-constriction cross over into propagating spin waves already at $\mu_0 H \sim$ 0.15 T, indicating a higher PMA than in the smaller nano-constriction. This is a general trend for different nano-constriction widths, which we believe is an effect of an etch induced reduction of PMA at the nano-constriction edges, which affects the smaller nano-constrictions in greater proportion. 

The field dependence in Fig.~\ref{fig:3} is entirely consistent with the expected behavior based on Fig.\ref{fig:1}b, where the influence of the PMA field strength on the non-linearity ($\mathcal{N}$) is depicted. In weak magnetic fields, $\mathcal{N}$ is negative, leading to localization of the auto-oscillations, but in stronger fields, $\mathcal{N}$ changes sign to positive, resulting in propagating spin waves. The strong PMA clearly allows one to achieve a positive $\mathcal{N}$ %the positive nonlinear frequency shift 
at a lower out-of-plane magnetization angle $\theta_{\text{M}}$, which in turn effectively results into a higher spin-Hall efficiency via sin $(90^{\circ}-\theta_{\text{M}})$ dependence and, therefore, substantially reduces the operational current\cite{giordano2016scirep}. With strong PMA field due to thinner CoFeB layer in the present case, the lowest threshold current density of auto-oscillations is obtained as 1.5 $\times$ 10$^{7}$ A/cm$^2$, which is considerably lower compared to the one observed in our recent study on relatively thicker CoFeB based W/CoFeB/MgO SHNOs with negligible PMA\cite{Zahedinejad2018apl}.

\begin{figure}
  \begin{center}
  \includegraphics[width=16cm]{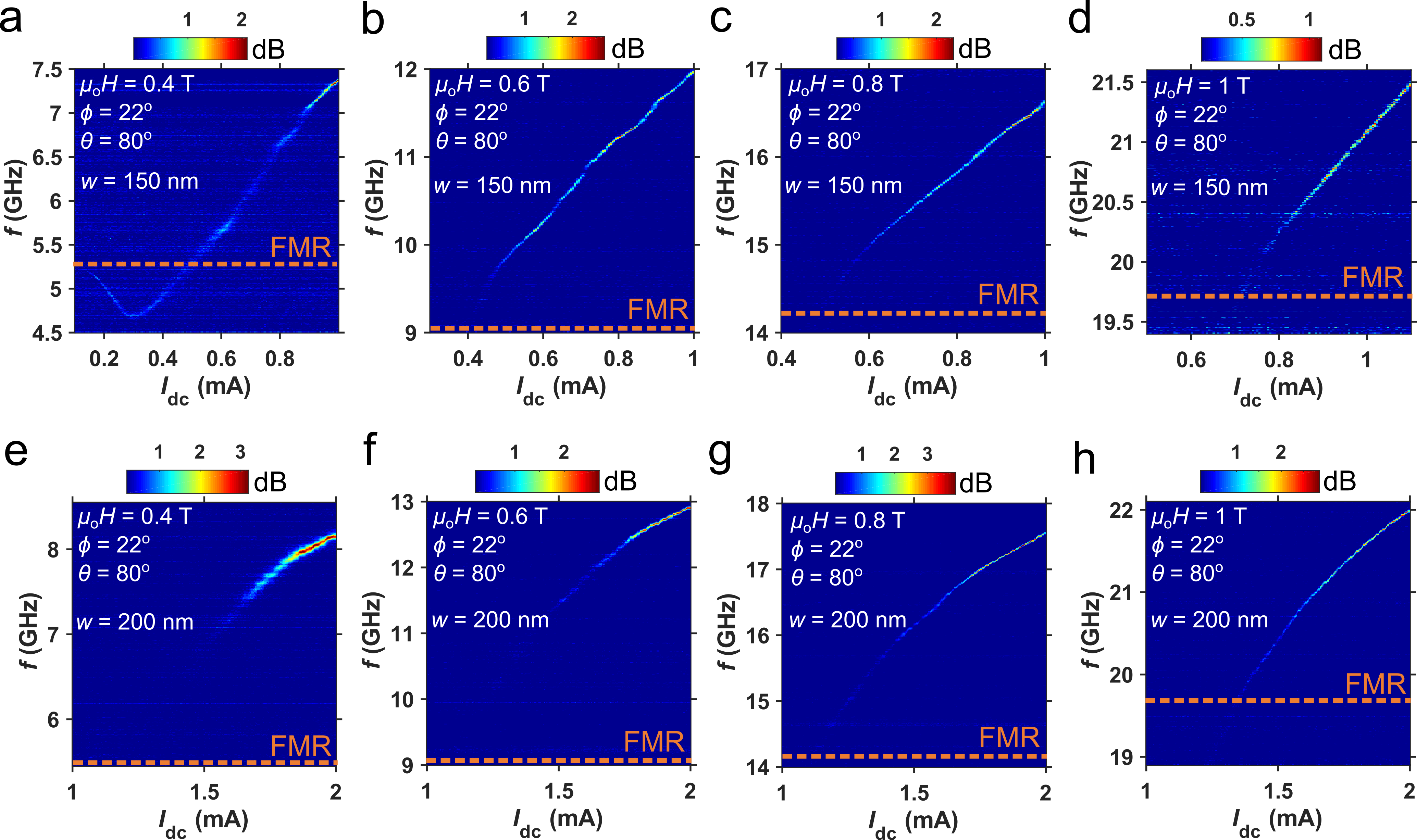}
    \caption{\textbf{Current tunability of spin wave auto-oscillations at different out-of-plane applied field strengths.} PSDs of the spin-wave auto-oscillations vs.~current for (a-d) a $w=$ 150 nm SHNO and (e-f) a $w=$ 200 nm SHNO subject to four different out-of-plane field strengths (0.4, 0.6, 0.8, and 1.0 T). Orange dashed lines indicate the FMR frequency. % at different field strengths under identical conditions.
    }
    \label{fig:4}
    \end{center}
\end{figure}

Having demonstrated the ability of nano-constriction SHNO devices to generate propagating spin waves, we now present the current tunability of propagating mode under fixed OOP field strengths. Fig.~\ref{fig:4}a-d show the current dependent PSD plots for a 150 nm wide SHNO. At 0.4 T, we observe the non-monotonic current dependence accompanying a red-shift in the auto-oscillation frequency at lower currents followed by the blue shift at higher currents. The FMR frequency, as shown by dashed line, distinguishes the two opposite non-linearity regimes observed in our device indicating a dramatic crossover from a localized behaviour of auto-oscillations at small currents to propagating one at higher currents. This peculiar non-linearity driven non-monotonic behaviour at lower fields is a manifestation of a gradual change in confinement potential of the auto-oscillation mode with current and is consistent with the theoretical predictions \cite{Dvornik2018prappl,Dvornik2018arxiv} and our simulation results discussed below. At higher fields, 0.6 T $<\mu_0 H<$ 1 T, only a blue-shifted behaviour %of frequency with increasing current 
is observed, highlighting the dominance of large positive $\mathcal{N}$ caused by the PMA. It is noteworthy that the variation of auto-oscillation frequency with electric current results in a very large positive value of the current tunability (${df}/{dI}$) reaching values over 4 GHz$/$mA in our devices (see Fig.~\ref{fig:4}b). The current dependence for the wider (200 nm) SHNO, shown in Fig.~\ref{fig:4}e-h, only exhibits a blue-shifted auto-oscillation frequency starting from the threshold current in all fields, 0.4 T $\leq \mu_0 H\leq$ 1 T. It is interesting to note that we no longer observe any auto-oscillation localization %of auto-oscillations 
in the wider nano-constriction, even at the lowest field of % while keeping the applied external field strength at 
0.4 T, which is consistent with stronger PMA in wider nano-constrictions. We also emphasize that the spectral linewidth of auto-oscillations $\Delta{f}<$ 20 MHz, extracted using a Lorentzian fit, %of auto-oscillation peak
yields a quality factor $Q={f}/\Delta{f}$ of up to 1000, indicating a considerably higher degree of oscillation coherence of the generated propagating spin-waves in our devices. In addition, our demonstration does not require very large values of PMA field to excite propagating SWs while the generation takes place in a wider frequency spectrum ranging from 5 to 22 GHz \cite{evelt2018pra}.

\subsection{Micromagnetic simulations}
\begin{figure}
  \begin{center}
  \includegraphics[width=16cm]{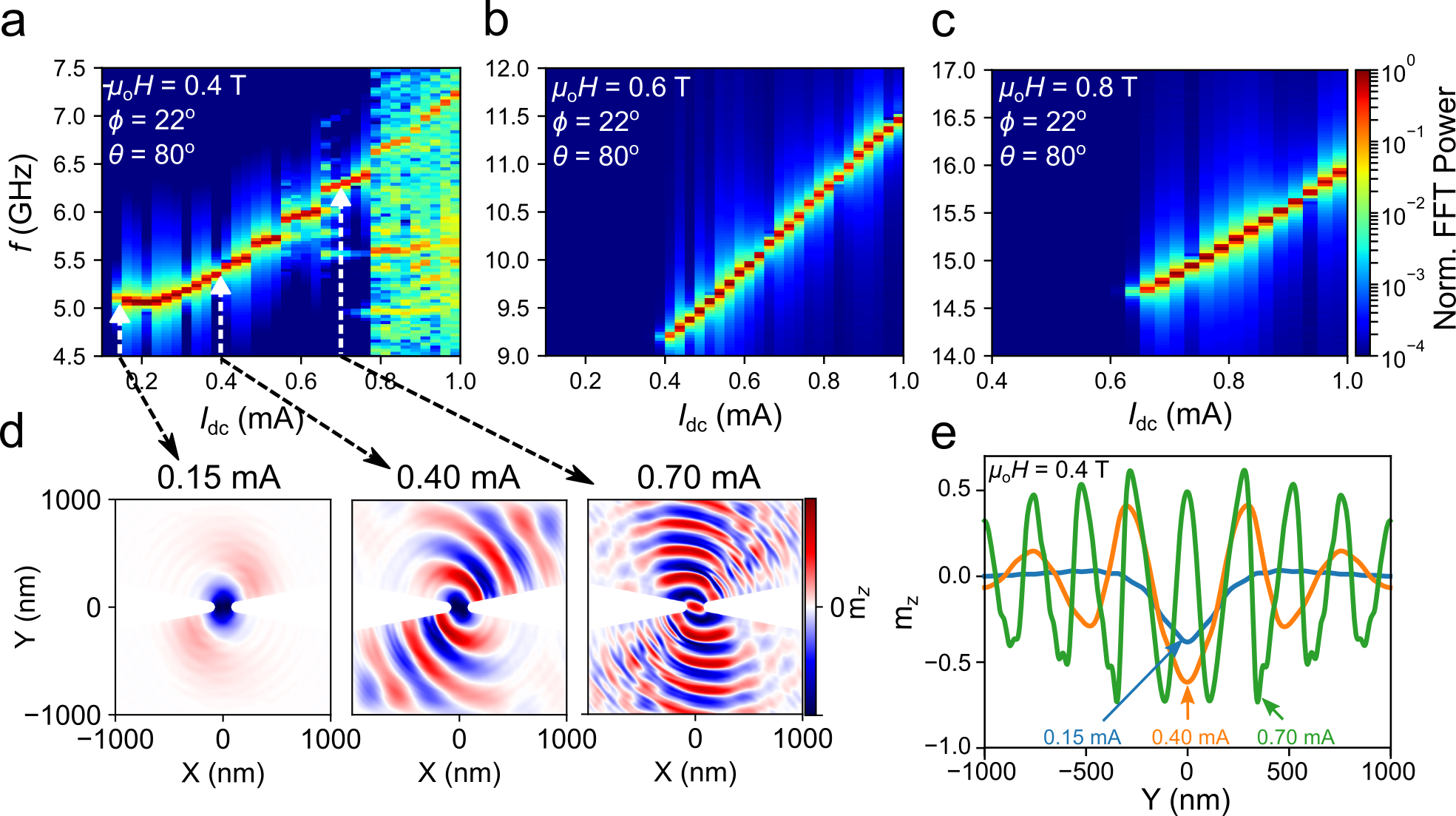}
    \caption{\textbf{Micromagnetic simulations.} (a-c) Micromagnetically simulated PSDs as a function of applied direct current through a 150 nm nano-constriction under % SHNO device for comparable external 
    field conditions used in the experiments. %employed in electrical measurements. 
    (d) Snapshots of the instantaneous $m_z$ component at three different currents showing how the auto-oscillations transition from localized (0.15 mA) to propagating (0.4 \& 0.7 mA) with a wave vector that increases with current.  (e) Cuts through Figure \ref{fig:5}d along the Y axis at the three different currents.}% The distribution of the mode energy at different simulated currents shows the gradual transition of localized auto-oscillations to propagating ones around the center of the nano-constriction.}
    \label{fig:5}
    \end{center}
\end{figure}

Finally, we present micromagnetic simulations performed using comparable conditions as in our electrical measurements to study the spatial profiles of the SW auto-oscillations for a 150 nm nano-constriction SHNO. All the magnetodynamical parameters used in the simulations are directly taken from the ST-FMR measurements discussed above. Fig.~\ref{fig:5}a-c show the current dependent PSD under three different fixed OOP field strengths indicating an excellent agreement with our experimental results in Fig.~\ref{fig:4}a-c. At 0.4 T, we observe a similar non-monotonic current dependence of the auto-oscillation frequency with simulated current, starting with a red-shifting frequency followed by a blue-shifted behavior (see Fig.~\ref{fig:5}a). It is interesting to note the appearance of multiple frequency steps at the lowest field. These are likely related to discreet changes in the wave vector and can also be observed in our experimental results at the same field (Fig.~\ref{fig:4}a), albeit as smoother transitions. At higher applied fields, these mode transitions disappear both in the experiments and in the simulations. The auto-oscillations then only show % as shown in Figure \ref{fig:5}b-c, simulated currents sweeps exhibit the blue-shifted auto-oscillation
a blue-shifted frequency behaviour. % in line with experimental results.

To gain deeper insight into %what constitutes 
the non-monotonic %auto-oscillation 
frequency behavior, we plot the spatial profiles of the simulated auto-oscillations %mode simulated 
at three representative currents in Fig.~\ref{fig:5}d. The nature of the auto-oscillations is qualitatively different at low and high current: in the low-current region, where $df/dI<0$, the auto-oscillations are clearly localized to the vicinity of the nano-constriction; at the two higher currents, where $df/dI>0$, the SWs clearly propagate with a wave vector that increases with current. Fig.~\ref{fig:5}e shows snapshots of % we compare the 
the instantaneous $m_z$ component at the three different currents, highlighting the transition from a localized nature %of auto-oscillations 
at 0.15 mA to propagating spin waves with about twice as large wave vector at 0.70 mA compared to 0.40 mA.

\section*{Discussion}
The capability to generate high frequency SOT-driven coherent propagating spin waves in metal based CMOS-compatible SHNO devices has particular potential for a number of reasons. First, the SW generation takes place already at very small operational currents with thresholds as low as 0.15 mA, with a straightforward development path towards even lower currents, making these oscillators the most amenable to adaptation in nanomagnonic circuits. Thanks to the large spin Hall angle provided by the $\beta$-W layer together with the strong PMA due to thinner CoFeB layer, the critical threshold current density required to excite propagating spin wave with SOT in metal devices has been reduced by about two orders of magnitude (10$^{7}$ A/cm$^2$) compared to theoretical predictions based on Pt and zero PMA\cite{giordano2014apl}. Further reduction of the critical current density should also be possible by yet higher PMA in yet thinner CoFeB. % layer and reducing the nano-constriction width. 
%In addition, the device layout flexibility and easier nano-fabrication process of these nano-constriction SHNOs allows the possibility to downscale nanomagnonic circuit elements. 
Second, %the effective anisotropy controllability exhibited by these SHNOs over highly rich non-linear magnetization dynamics serves to bridge the link between spin-orbitronics and nanomagnonics\cite{divinskiy2018arxiv}. The 
their current tunability should allow these SHNO to rapidly switch between localized and propagating spin waves, effectively acting as tunable nanoscale sources of ultra-short SW pulses and wave packets\cite{divinskiy2016apl}. %Third, the CMOS compatibility exhibited by these SHNOs further adds considerable weight to their rich technological significance by enabling monolithic integration into the existing CMOS circuits.
Thirdly, the wave nature of the propagating SWs will now allow for experimental realizations of SOT driven spin wave computing \cite{klingler2014apl,fischer2017apl} such as interference based majority gates.  %minimal fabrication complexity of our devices will enable us to capitalize these 
Finally, the propagating spin wave modes could further boost the long-range mutual synchronization of SHNO chains and networks %of large number of SHNOs with substantial electrical read-out 
for neuromorphic computing applications. Given the recent experimental demonstration of neuromorphic vowel recognition using four mutually synchronized STNOs\cite{romera2018nature}, we believe that the long range mutual synchronization driven by propagating spin waves in low operational current SHNOs may turn out to be the most viable solution for scaling neuromorphic computing to large dynamical neural networks.

\begin{methods}
\subsection{Nano-constriction device fabrication.}
A trilayer stack of W(5)/Co$_{20}$Fe$_{60}$B$_{20}$(1.4)/MgO(2) (thicknesses in nm) was grown at room temperature on an intrinsic high resistivity Si substrate ($\rho_{Si}>$10 k$\Omega \cdot$cm) using an AJA Orion-8 magnetron sputtering system. DC and RF sputtering were sequentially employed for the depositions of metallic and insulting layers, respectively. The chamber was evacuated to a base pressure of 2$\times$10$^{-8}$ mTorr while the Ar gas pressure was maintained at 3~mTorr during the growth of all the layers. The deposition rate of W was kept at 0.09~\AA/s to obtain high resistivity $\beta$-phase exhibiting a large spin Hall angle \cite{mazraati2016apl,zhang2016apl,Zahedinejad2018apl}. The same deposition rate was maintained for Co$_{20}$Fe$_{60}$B$_{20}$ layer, while MgO layer was grown at 0.04~\AA/s. The \textit{as-deposited} stack was subsequently annealed under chamber base pressure at 300~$^{\circ}$C for 1 hour to induce PMA. The annealing process was followed by deposition of 4 nm of SiO$_x$ to protect the MgO layer from degradation due to exposure to the ambient conditions. The resistivities of $\beta$-phase W and Co$_{20}$Fe$_{60}$B$_{20}$ were measured as 213 $\mu \Omega \cdot$cm and 100 $\mu \Omega \cdot$cm, respectively. 
The trilayer stack was then patterned into an array of 4$\times$14$~\mu m^2$ rectangular mesas and the nano-constriction SHNO devices with different widths were defined at the center of these mesas by a combination of electron beam lithography and Argon ion beam etching using negative electron beam resist as the etching mask. In addition, we patterned 6$\times$18$~\mu m^2$  microstrips for ST-FMR measurements. The fabrication process is detailed in Ref. \citen{Zahedinejad2018apl}.

\subsection{Analytical calculation of non-linearity coefficient for a thin magnetic film.}
To calculate the nonlinear coefficient $\mathcal{N}$, shown on Figure \ref{fig:1}b, we start from the magnetic energy density, which consists of Zeeman, dipolar and PMA terms. Employing a well-known method \cite{slavini2005approximate,slavin2009nonlinear}, in which the Hamiltonian is expressed in the elliptically polarized dimensionless variables by a sequentially applying Holstein-Primakoff and Bogolyubov transformations with the further elimination of the non-resonant three-waves processes, one can derive the final result (see Ref. \citen{mohseni2018prb} for the details) as:
\begin{equation}
    N=\frac{2 \omega_{0}}{\mathcal{A}}\left(\mathcal{T}-3\frac{|\mathcal{W}_1|^2+|\mathcal{W}_2|^2}{\omega_0}\right),
    \label{eq:nonlinearity}
\end{equation}
where
\begin{align*}
    \mathcal{T} &=\left[3(u^2+|v|^2)^2-1\right]\mathcal{U}_1/2-3u (u^2+|v|^2)\left(v\mathcal{U}_2+v^*\mathcal{U}_2^*\right), \\
    \mathcal{W}_1 &= 3(u^2+|v|^2)(u\mathcal{V}-v^*\mathcal{V}^*)/2-(u\mathcal{V}+v^*\mathcal{V}^*)/2, \\
     \mathcal{W}_2 &= - u v^*(u\mathcal{V}-v^*\mathcal{V}^*).
\end{align*}
The FMR frequency, shown by a solid line on Figure 3, can be calculated as:
\begin{equation}
    \omega_{0}=\sqrt{\mathcal{A}^2-|\mathcal{B}|^2}.
    \label{eq:FMRfreq}
\end{equation}
The coefficients of Bogolyubov transformation are:
\begin{equation*}
    u=\text{sign} (\mathcal{A}) \sqrt{\frac{\mathcal{A}+\omega_0}{2 \omega_0}}, \qquad v=\frac{\mathcal{B}^*}{|\mathcal{B}|}\sqrt{\frac{\mathcal{A}-\omega_0}{2 \omega_0}},
\end{equation*}
where
\begin{align*}
    \mathcal{A} &= \omega_{\text{H}}-\frac{1}{2}(\omega_{\text{k}}-\omega_{\text{M}})\cos^2 \theta_{\text{M}},\\
    \mathcal{B} &= -\frac{1}{2}(\omega_{\text{k}}-\omega_{\text{M}})\cos^2 \theta_{\text{M}},\\
    \mathcal{V} &= (\omega_{\text{M}}-\omega_{\text{k}})\sin \theta_{\text{M}} \cos \theta_{\text{M}}, \\
    \mathcal{U}_1 &= (\omega_{\text{k}}-\omega_{\text{M}})\sin \theta_{\text{M}} \left(\frac{3}{2} \cos^2 \theta_{\text{M}}-1\right), \\
    \mathcal{U}_2 &=-\mathcal{B}/2.
\end{align*}
In the above expressions we used the notations $\omega_{\text{M}}= \gamma \mu_0 M_{\text{S}}$, $\omega_{\text{k}}=\gamma \mu_0 H_{\text{k}}^{\perp}$, $\omega_{\text{H}}=\gamma \mu_0 H_{\text{M}}$, where $M_{\text{S}}$ is the saturation magnetization, $H_{\text{k}}^{\perp}$ is the PMA field, $H_{\text{M}}$ -- effective internal field. The latter can be defined together with the internal angle of magnetization $\theta_{\text{M}}$ from the following equations:
\begin{align*}
H_{\text{M}} \cos \theta_{\text{M}} &= H_{\text{ex}} \cos \theta_{\text{ex}} \\
(H_{\text{M}}-H_{\text{k}}^{\perp}+M_{\text{S}}) \sin \theta_{\text{M}} &= H_{\text{ex}} \sin \theta_{\text{ex}},
\end{align*}
where $H_{\text{ex}}$ is the externally applied magnetic field applied at the out-of-plane angle $\theta_{\text{ex}}$.

\subsection{ST-FMR measurements.}
The magneto-dynamical parameters of devices under investigation were determined by performing ST-FMR measurements at room temperature on a 6$\times$18$~\mu m^2$ wide microstrip of W(5)/Co$_{20}$Fe$_{60}$B$_{20}$(1.4)/MgO(2) stack. A radio-frequency (RF) current modulated at 98.76 Hz is made to inject through a high frequency bias-T through the microstrip at a frequency characteristic of FMR (3-12 GHz), generating spin-orbit torques as well as Oersted field under an external applied IP magnetic field. The resulting torques excite the magnetic moment in the CoFeB layer to precess leading to a time-dependent change in the resistance of the microstrip due to the AMR of CoFeB layer\cite{liu2011prl}. The oscillating AMR mixes with the RF current to create a d.c. voltage, $V_{\text{mix}}$, across the microstrip and is measured using the circuit displayed in inset of Fig. \ref{fig:2}. All ST-FMR measurements shown in Fig. \ref{fig:2} were carried out by sweeping an in-plane field ($\phi$ = 30$^{\circ}$) from 350 to 0 mT, while the frequency of the input RF signal is kept fixed. To determine the SHA, we injected small dc currents in addition to RF current through \textit{dc} and \textit{rf} ports, respectively of a bias-T. The resonance feature in voltage response from each field sweep was fitted to a sum of one symmetric and one anti-symmetric Lorentzian sharing the same resonance field and linewidth. In Fig. \ref{fig:3}, ST-FMR measurements on the microstrip were performed by sweeping the applied field at a fixed IP angle of $\phi$ = 22$^{\circ}$ and OOP angle of $\theta$ = 80$^{\circ}$ to measure FMR frequency under the identical conditions employed during auto-oscillation measurements.

\subsection{Microwave measurements.}
All microwave electrical measurements were carried out at room temperature using a custom built probe station with the sample mounted at a fixed in-plane angle on an out-of-plane rotatable sample holder between the pole pieces of an electromagnet capable of producing a uniform magnetic field. A direct positive electric current, $I_\text{dc}$, was made to inject through \textit{dc} port of a high frequency bias-T and the resulting auto-oscillating signal was then amplified by a low-noise amplifier with a gain of $\geq$ 32 dB and subsequently recorded using a spectrum analyzer from Rhode \& Schwarz (10 Hz-40 GHz) comprising a low resolution bandwidth of 300 kHz. We measured multiple SHNO devices and restricted the maximum current up to 1 mA for 150 nm and 2 mA for 200 nm nano-constriction devices in order to avoid irreversible changes due to device degradation in the output microwave characteristics.

\subsection{Micromagnetic simulations.}
The micromagnetic simulations were performed using the Graphics processor Unit (GPU)-accelerated program MUMAX3\cite{Vansteenkiste2014} with input provided by the COMSOL simulations. A 150 nm nano-constriction width SHNO device is modelled into $1024\times1024\times1$ cells with an individual cell size of $3.9\times3.9\times5$ nm$^3$. The material parameters employed in simulations such as the saturation magnetization $\mu_0 M_{\text{S}} =$ 0.93 T, the gyromagnetic ratio $\gamma/2\pi$ = 29.9~GHz/T, and the Gilbert damping constant $\alpha$ = 0.023 were obtained from STFMR measurements on a 6$\times$18$~\mu m^2$ wide microstrip of W(5 nm)/Co$_{20}$Fe$_{60}$B$_{20}$(1.4 nm)/MgO(2 nm) stack. We used PMA field of $\mu_0 H_{\text{k}}^{\perp}$= 0.57 T, slightly lower than the measured value of 0.62 T on a microstrip. The exchange stiffness constant of $A_{\text{ex}} = 19\times10^{-12}$ J/m for CoFeB was taken from Ref. \citen{sato2012cofeb}. The distribution of charge current density and the resulting Oersted field landscape for a W/CoFeB bilayer is obtained with COMSOL simulations. The corresponding spin current is then estimated from the simulated charge current in the W layer generating a transverse pure spin current along the out of plane direction while considering a spin Hall angle, $\mathit{\theta_{\text{SH}}}=$ -0.41, obtained from STFMR measurements on a 6$\times$18$~\mu m^2$ wide microstrip. The magnetization dynamics is simulated by integrating the Landau-Lifshitz-Gilbert-Slonczewski equation over 62.5 ns, with the first 31 ns discarded in the following analysis to exclude transient effects. The auto-oscillation spectra is obtained by performing the fast Fourier transform (FFT) of the simulated time evolution of the magnetization averaged over sample volume. The full spatial maps of the magnetization are extracted from the time domain data at a fixed time of 32 ns. 

%Put methods in here.  If you are going to subsection it, use
%\verb|\subsection| commands.  Methods section should be less than
%800 words and if it is less than 200 words, it can be incorporated
%into the main text.

%\subsection{Method subsection.}

%Here is a description of a specific method used.  Note that the
%subsection heading ends with a full stop (period) and that the
%command is \verb|\subsection{}| not \verb|\subsection*{}|.

\end{methods}

%% Put the bibliography here, most people will use BiBTeX in
%% which case the environment below should be replaced with
%% the \bibliography{} command.

% \begin{thebibliography}{1}
% \bibitem{dummy} Articles are restricted to 50 references, Letters
% to 30.
% \bibitem{dummyb} No compound references -- only one source per
% reference.
% \end{thebibliography}

%% Here is the endmatter stuff: Supplementary Info, etc.
%% Use \item's to separate, default label is "Acknowledgements"

\begin{addendum}
 \item This work was supported by the European Research Council (ERC) under the European Community’s Seventh Framework Programme (FP/2007-2013)/ERC Grant 307144 “MUSTANG.” This work was also supported by the Swedish Research Council (VR), the Swedish Foundation for Strategic Research (SSF), and the Knut and Alice Wallenberg Foundation.
  \item[Author contributions] H.F. carried out all the electrical measurements and their analysis.  M.Z. developed the material stacks and fabricated the devices.  R.K. performed the analytic calculations. M.D. performed the micromagnetic simulations.  J.\AA. initiated and supervised the project. All authors contributed to the data analysis and co-wrote the manuscript.
  \item[Competing Interests] The authors declare that they have no
competing financial interests.
 \item[Correspondence] Correspondence and requests for materials
should be addressed to J. \AA kerman~(email: johan.akerman@physics.gu.se).
\end{addendum}

\end{document}